\documentclass[%
reprint,
 amsmath,amssymb,
 aps,
]{revtex4-2}

\usepackage{graphicx}
\usepackage{dcolumn}
\usepackage{bm}

\newcommand{\npythia}{3,537}
\newcommand{\ncominatorics}{19,497}
\newcommand{\nkinematics}{159,674}

\newcommand{\pythia} {\textsc{Pythia8}}

\onecolumngrid

\begin{document}

\preprint{HEP-ANL-186383}

\title{Estimation of the chances to find  new phenomena  at the LHC in a model-agnostic combinatorial analysis}

\author{S/V.Chekanov}
 \altaffiliation{HEP Division, Argonne National Laboratory,
9700 S.~Cass Avenue, Argonne, IL 60439, USA.}

\date{\today}

\begin{abstract}
In this paper, we estimate the number of event topologies that have the potential to be produced in $pp$ collisions at the Large Hadron Collider (LHC) without violating kinematic and other constraints. We use numerical calculations and combinatorics, guided by large-scale Monte Carlo simulations of Standard Model (SM) processes. Then, we set the upper limit on the probability that new physics may escape detection, assuming a model-agnostic approach. The calculated probability is unexpectedly large, and the fact that the LHC has not found new physics until now is not entirely surprising. We argue that the optimal direction for maximizing the chances of finding new physics is to use unsupervised machine learning for anomaly detection or algorithms designed for event classification.
\end{abstract}


\maketitle


\section{Introduction}
\label{introduction}

A simple answer to the question of why the Large Hadron Collider (LHC) did not discover new physics can be rather short: there are no new TeV-scale phenomena that can be discovered using the LHC data, and perhaps this might be the most obvious outcome of the LHC for many years to come. Another way to answer this question is to argue that not all event classes have been explored so far. Physics beyond the Standard Model (BSM) can still produce unusual events with small production cross sections, but such events have not been found yet. Many make this last argument, but putting this discussion on solid quantitative footing requires removing the constraints imposed on experimental research by model builders. In the past, the LHC had too significant a focus on setting limits on proposed BSM models. Looking at this problem from a much wider angle by adopting a model-agnostic view of BSM searches could reveal many unexpected features of the LHC data.

The goal of this paper is to calculate the number of unique event classes produced at the LHC. We define an exclusive event class (or event topology) as a group of events with exactly the same number of identified particles and reconstructed jets. We use numerical calculations based on combinatorics, guided by large-scale Monte Carlo (MC) simulations of the Standard Model (SM). The latter sets the necessary kinematic constraints and boundary conditions for our calculations. Then we estimate the chances that new phenomena might have escaped detection at the LHC. These calculations are fully model-agnostic since they are not guided by BSM models.

\section{Number of unique topologies from combinatorics}

Let us estimate how many unique event topologies are expected to be produced in $pp$ collisions
at $\sqrt{s}=13$~TeV. Assume LHC collisions produce events with light-flavor 
jets ($j$), jets associated with $b$-quarks ($b$-jets), electrons ($e$), muons ($\mu$), tau ($\tau$) leptons and photons ($\gamma$). In addition, neutrinos can lead to missing transverse energy, referred to by the acronym ``MET''  (denoted in the numerical calculation by the letter $m$). 
All events can be grouped into exclusive classes denoted as 
$$
(Nm,\> Nj,\> Nb,\> Ne,\> N\mu,\> N\tau,\> N\gamma), 
$$
where $N$ is an integer number that defines the number of objects of a certain type, $v=j, b, e, \mu, \tau, \gamma$, in a collision event. In the following, the word ``object'' will be used for jets, $b$-jets, leptons, and photons.  
In the case of MET, $Nm$ is either $0m$ (no significant MET) or  $1m$ (when MET is above 200~GeV). Thus, an event class marked with $1m$ corresponds to  one (or several) produced neutrinos.  Using this notation, $(1m,\> 2j,\> 1b,\> 0e,\> 0\mu,\> 0\tau,\> 1\gamma)$ represents a class of events with large MET ($1m$), two light-flavor jets ($2j$), one $b-$jet ($1b$) and one photon ($1\gamma$).

To estimate how many unique topologies are expected from the SM, we have produced a sample of \pythia\> version  8.307 \cite{Sjostrand:2007gs} MC events with $pp$ collisions at $\sqrt{s}=13$~TeV after enabling all SM processes of this generator. Similar to \cite{Chekanov:2017pnx}, the simulation used 44 physics sub-processes 
at leading-order QCD, such as light-flavor dijet production, all top production, weak single and double boson production, prompt photons and all Higgs
SM processes. The cut on the two-body matrix elements in \pythia\> was set to 100~GeV. The total integrated luminosity of the simulation was 154~fb$^{-1}$, i.e. larger than the LHC Run2 data sample of 140~fb $^{-1}$. Since light-flavor QCD dijets cannot create the required complexity of the event classes, the generation of such events was relatively suppressed compared to other sub-processes with lower cross sections. 
The total number of generated events was 0.53 billion.  Stable particles with a lifetime larger than $3\cdot 10^{-10}$ seconds were considered, while neutrinos were excluded from consideration. 
The NNPDF 2.3 LO~\cite{Ball:2014uwa} parton density function, interfaced with \pythia\> via the LHAPDF library~\cite{Buckley:2014ana}, is used. 
A detector simulation was not applied. The object reconstruction was the same as in \cite{Chekanov:2017pnx}.
Hadronic jets are reconstructed using the anti-$k_T$ algorithm~\cite{Cacciari:2008gp} with a distance parameter of $R=0.4$ implemented in the {\sc FastJet} package~\cite{Cacciari:2011ma}.
The transverse momenta ($p_T$) of the jets must be greater than $20$~GeV, and the pseudorapidity ($\eta$) must satisfy $|\eta|<2.5$.
A jet is classified as a $b$-jet if its four-momentum matches the momentum of a $b$-quark, and the $b$-quark contributes more than 50\% of the total jet energy. 
Leptons and photons are required to be isolated.
A cone of size 0.2 in azimuthal angle ($\phi$) and $\eta$ is deﬁned around the true
direction of the lepton. Then, all energies of particles inside this cone were summed up. 
A lepton is considered isolated if it carries more than 90\% of the cone energy.
The transverse momentum cut and the $\eta$ cut were the same as
for jets.  The requirement for MET was 200 GeV, i.e. when $0m$ becomes $1m$ in the symbolic calculation. 

According to the above SM MC simulation,  the number of non-identical event topologies was \npythia. The maximum number of observed objects was $17j$, $8b$, $4e$, $4\mu$, $4\tau$ and $4\gamma$.  We did not observe more than 20 objects per event. 
The total number of light-flavor $b-$jets was always less than 19.  In addition, the total number of leptons 
was never larger than 5. All such restrictions can be called the ``boundary'' condition, which limits the number of possible event classes. They are summarized below:
\begin{equation}
 \begin{array}{l}
Nm<2, \quad Nj<18, \quad  Nb<9, \quad Nj+Nb<19, \\
Ne<5, \quad N\mu<5, \quad N\tau<5, \quad  N\gamma<5, \\
Nj+N\ell<19,\quad Nj+N\gamma<19, \\
Nb+N\ell<9, \quad Nb+N\gamma<9, \\
N\ell<6, \quad  N\ell+N\gamma<6, \\
Nj+Nb+N\ell+N\gamma<21,
\end{array}
\label{boundary}
\end{equation}
where $N\ell=Ne+N\mu+N\tau$ is the total number of leptons.
Events with $9j$ and  $5b$  (but with transverse momenta larger than $p_T>20$~GeV used in this paper) and four-lepton  events have been recently studied by the LHC \cite{ATLAS:2020wgq,CMS:2019hsm}, thus our MC simulation should be a reasonable  representation of the reality.

We will keep a conservative view that a BSM phenomenon does not violate the boundary condition;
otherwise, it can easily be found by looking at inclusive single-particle distributions of identified particles or jets. For example, an observation of events with five muons could alarm the observer in the past and thus such high-multiplicity events cannot represent the experimental 
challenge for their detection.   
But new phenomena can be ``hidden'' in exclusive combinatorial combinations, which are more intricate to discover experimentally.
We will come back to the discussion of this point later. 

Let us calculate how many combinatorial combinations are expected by preserving the SM boundary condition Eq.~(\ref{boundary}). First, we set the maximum number
of objects to be observed to $Nmax=20$. There are up to $n=7$ objects per event (where MET is counted as an additional ``object''). 
The total number of combinations, where items can be repeated more than once and the ordering of items is not important, is
\begin{equation}
\sum_{r=2}^{Nmax} \frac{ (n+r-1)! } { (n-1)! r!}.  
\end{equation}
Thus, the total number of unique combinations is 888022.  Imposing the boundary condition Eq.~(\ref{boundary}) from the SM MC simulation is not straightforward using analytic calculation. However, such a calculation can be obtained numerically as shown in Appendix~\ref{code}.  
The obtained answer is \ncominatorics~combinations.

The difference between the MC prediction (\npythia) for SM processes and what, potentially, can be expected for the number of event classes from combinatorics (\ncominatorics) 
demonstrates that the MC event sample does not include all possible event classes.  For example, event topologies such as

\begin{equation}
\begin{split}
    (0m,\> 2j,\> 2b,\> 2e,\> 3\mu,\> 0\tau,\> 0\gamma),  
    \\
    (1m,\> 4j,\> 0b,\> 3e,\> 1\mu,\> 1\tau,\> 1\gamma), \\
    etc.
    \label{example}
\end{split}
\end{equation}
have never been seen in the generated event sample for SM processes. The author does not know which BSM scenario can lead to such event classes.
Note that the event topologies are defined in the
restricted phase space, i.e. in the limited kinematic region defined by the transverse momentum and pseudorapidity selection.
Thus, such event topologies cannot violate charge, lepton number, energy-momentum conservation, and other constraints.
The difference of about 5 between the number of event classes predicted by the \pythia\> generator  and by the numeric combinatorics can be an indication that the event generation may require more events. In addition, not all physics processes are 
included in the event generator.  
For example, next-to-leading  order QCD effects may be in play. It should be noted that \pythia\> agrees 
well with alternative MC simulations up to six jets \cite{ATLAS:2019mra}, but the other event topologies need to be verified too.   
We will put this question aside and assume that the total number of possible event classes is \ncominatorics\>, as derived in the numeric computation  with the \pythia\> boundary condition Eq.~(\ref{boundary}), but not what has been predicted by \pythia\> itself for the number of event classes.
More realistic event generators may reduce the discrepancy between our numeric estimate and the generator predictions for the number of unique event classes, but they cannot change the conclusion of this paper, which does not rely on MC simulations.

How can we be sure that previous LHC studies were able to explore all such event topologies? 
According to the publication record of the ATLAS and CMS experiments,  $pp$ collision 
events have been studied in about 600 publicly available results using 140~fb$^{-1}$ of data.   
For the sake of argument, let us assume that 5 non-identical event classes
were scrutinized in each paper\footnote{In our view, this number of 
studied event classes is very optimistic. Typically, an experimental paper has only a few such distributions, and most of them are too inclusive to pinpoint a specific event class. In addition, ATLAS and CMS often repeat measurements using similar final states. Due to the lack of exact statistics on published distributions, we prefer to consider the most optimistic scenario.}, 
and they were found to match the SM predictions.
This gives 3,000 investigated event classes.  Note that one ATLAS publication \cite{ATLAS:2018zdn} contains the studies of more than 700 event classes, but that analysis used a small fraction of the LHC Run2 data, and these event classes are
expected to be a subset of the 3,000 event topologies assumed before.
Therefore, the number of unexplored event topologies, out of \ncominatorics~expected,  
is  close to $81\%$.
 
If one considers fully reconstructed (identified) SM heavy particles, such as $Z$, $W$ and top quarks,
the number of event classes will increase. This can easily be checked by adding these additional particles in our numeric notation after reducing the maximum multiplicities of leptons from 4 to 3 (i.e. considering $W\to \ell\nu$ decays) and reducing the number of jets ($b$-jets) by one. We should also require that the total number of $W$, $Z$ and top quarks cannot be greater than 3; the latter boundary condition makes this example more realistic. In this case, the number of event topologies will increase to more than 140,000.

\section{Discussion}

When discussing the coverage of the event classes by the LHC studies, it is assumed
that new phenomena predominantly contribute to a single event class, 
rather than to many event classes. The latter assumption, keeping in mind our model-agnostic approach, 
should be quite reasonable considering the fact that we do not know much about what can be expected from BSM physics. 

From the standpoint of QCD, even if a BSM model is characterized by a very specific event class
(say, with a fixed $nj$ number jets plus some fixed number of leptons and photons), additional event classes
with extra jets can be produced due to the parton shower. The event rate of the events with $nj$+1 jets is suppressed with respect to events with $nj$ jets by the strong coupling constant $\alpha_S$ (times the number of jets). 
However, ignoring softer jets involves an additional supposition that there is nothing interesting in high-jet multiplicity events since they originate from the QCD parton showering. This assumption is incompatible with the general search strategy since it must involve an undefined cutoff parameter that 
limits the number of jets and the entire scope of model-agnostic searches.
This is why an exclusive approach to jet multiplicity has been adopted by the ATLAS \cite{ATLAS:2018zdn} general searches. 

From  an experimental perspective, it is not unreasonable to think that some inclusive measurements may have 
certain sensitivity to the \ncominatorics\> event classes 
reported for the condition Eq.~(\ref{boundary}).
This is because many studied distributions at the LHC are a ``mix'' of different event classes.
In our view, inclusive measurements cannot effectively 
pinpoint a specific event topology produced with a small cross-section. Generally, searches in events with exclusive definitions of jets and particles, where any event class with a fixed jet multiplicity is treated as a unique hadronic-final  signature, are better motivated. 
For example, it is difficult to understand how an inclusive 2-jet measurement can ping point event class with additional 2 jets and a few leptons shown in Eq.~(\ref{example}), which may have a cross-section by several orders of magnitude smaller than the inclusive 2-jet measurements.
Thus, it is necessary to carry out dedicated measurements focusing on such exclusive event topologies.

\section{Experimental scenario}
\label{exper}
\begin{figure*}[ht]
  \begin{center}
  \includegraphics[width=0.7\textwidth]{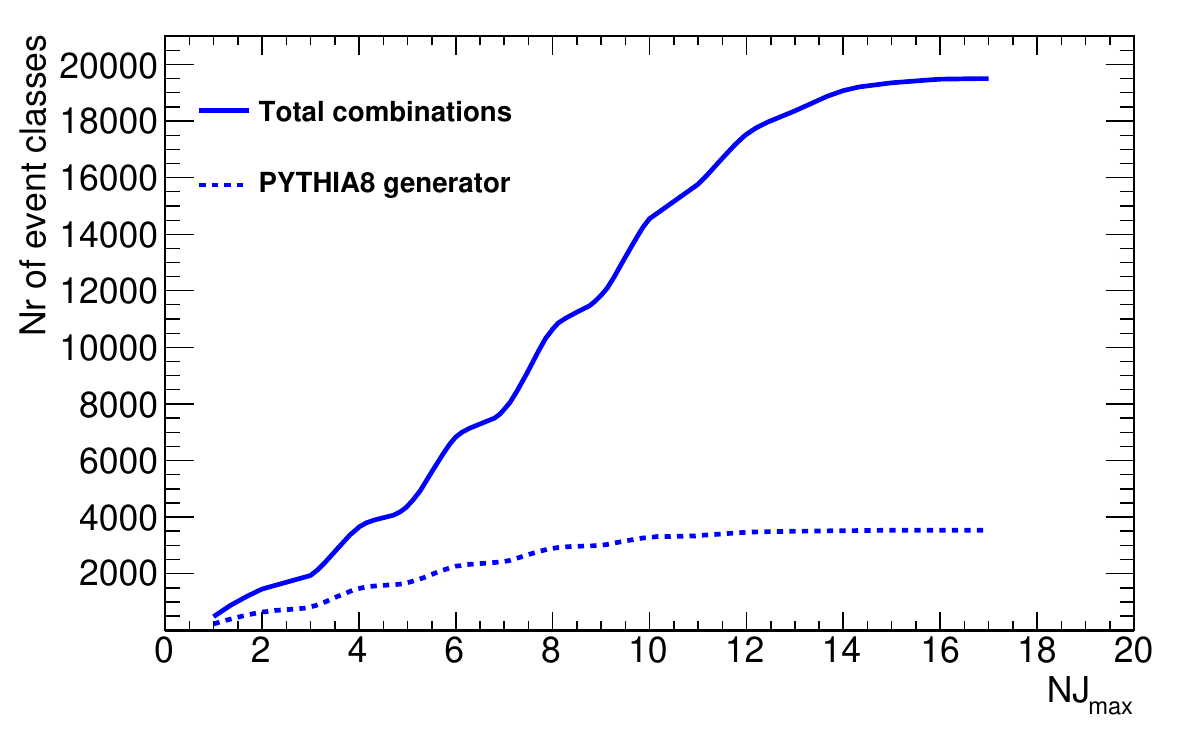}
  \end{center}
  \caption{
   The number of expected event classes from the combinatorial analysis and \pythia\> as a function of the maximum number of jets ($\mathrm{NJ}_{max}$). It is assumed that the number of $b-$jet should be less than 50\% of the total number of jets.
  }
  \label{fig:multiplot}
\end{figure*}

We also verified a scenario after changing the \pythia\> boundary condition Eq.~(\ref{boundary}) 
to:
\begin{equation}
 \begin{array}{l}
Nm<2, \quad Nj<7, \quad  Nb<5, \quad Nj+Nb<10, \\
Ne<4, \quad N\mu<4, \quad N\tau<4, \quad  N\gamma<4, \\
N\ell<5, \quad  N\ell+N\gamma<5, \\
Nj+Nb+N\ell+N\gamma<13.
\end{array}
\label{boundaryH}
\end{equation}

These restrictions are within the reach of the current LHC studies since they correspond to the realistic reconstruction of multiple hadronic jets \cite{ATLAS:2020wgq,CMS:2019hsm}. The obtained number of event classes from the numeric calculation is 6,676. \pythia\> predicts 2,485 event classes, thus the condition in Eq.~(\ref{boundaryH}) reduces the discrepancy between our expectation and the generator. Assuming 3,000 investigated event classes at the LHC, the number of unexplored event topologies is about 55\%.

The trigger thresholds used at the LHC are another noteworthy point for consideration. The minimum cut on the transverse momenta $p_T > 20-30$ GeV of jets, which can lead to large jet multiplicities, is rather low for effective trigger selection and for high purity in the reconstruction of jets. However, it should be pointed out that the main triggers to be used in such searches do not need to be based on jet triggers: Since almost every event contains $e$, $\mu$, or $\gamma$, the events can effectively be triggered by using these particles, for which a $20$ GeV requirement is not unusual. From the point of view of triggers, the most difficult categories for detection are the events with very few jets without associated production of electromagnetic particles or significant MET. But the number of such classes is very small compared to the overall number of combinations, thus it should not change much our conclusion regarding the total number of expected event classes.

\section{Hard-QCD scenarios}

The previous consideration deals with situations when the numbers of jets can go up to 6.
This can be a realistic scenario for some supersymmetric extensions of the Standard Model, where jets stem from exotic particles.
However, many BSM models can only be characterized with several hard-$p_T$ jets (say, up to 4), while the other jets are produced due to the parton shower, where the event rate of the events with $Nj$+1 jets is suppressed with respect to the events with $Nj$ jets by the strong coupling constant $\alpha_S$ (times the number of jets). 

Therefore, we will consider a more
inclusive definition of event classes using Eq.~\ref{boundaryH}, but with the $Nj<5$ and $Nb<3$ restrictions. 
Electromagnetic particles do not shower; therefore, we preserve the original boundary condition for $e$, $\mu$, $\tau$ and $\gamma$.
This implies that we ignore events with higher jet multiplicities since
they originate from the parton shower of quarks and gluons (but $\alpha_S$ suppressed) of the same process, thus they do not carry the new information about such BSM events. 
The obtained number of event classes from the numeric calculation is 3172.
About 40\% of such classes were not found in \pythia.
Such a strong reduction in event classes compared to events with larger jet multiplicities 
may also lead to conclusion that many of such events have been studied at the LHC.

Figure~\ref{fig:multiplot} demonstrates the dependence of the number of possible event classes
as a function of the maximum number of jets ($\mathrm{NJ}_{max}$).
The calculations are presented for 
the numerical combinatorial analysis and for the \pythia\> simulation.
During the calculation, we limit the number of $b-$jets to 50\% of the $NJ_{max}$ value. It can be concluded that the discrepancies between the expected number of event classes and \pythia\> become smaller for small 
values of $\mathrm{NJ}_{max}$.

\section{High-statistics scenario}
\label{high_stat}

Now we will come back to the original condition Eq.~(\ref{boundary}), and ask the following question: 
what is the number of classes with more than 9
events. This would correspond to about $3\sigma$ evidence for an observation of such event class. This will address the low-statistics problem since this requirement removes low-statistics events, which do not have the potential to lead to ''evidences'' for experimental observations.

Such estimates cannot be done in a model-independent way since we do  not have the information about cross-sections of event classes which are not in \pythia\>  (or any Monte Carlo simulation).
The obtained number of combinations in  \pythia\> with more than 9 events is 1,958 (out of the total \npythia). 
Thus, the total number of events which have too low statistics for any observational 
evidence is about 45\%.

In reality, we cannot perform such estimates for hypothetical BSM processes with unknown cross sections. But if we assume that the same fraction of low-statistics events holds for \ncominatorics\> classes using 
the condition Eq.~(\ref{boundary}), we arrive to the number 10,723
for event classes with more than 9 events in each. Assuming 3,000 investigated event classes at the LHC,
the number of unexplored event topologies is about 72\%. Of course, this number is very speculative
since we do not have any prior knowledge of the cross-sections of such hypothetical BSM processes. 

\section{Kinematic consideration}

It is more difficult to understand the kinematic side of the argument beyond the 
object-multiplicity combinatorics.
So far we assumed that all objects are produced in any detector region, following some
density distributions expressed
in terms of $p_T$, $\eta$ and $\phi$, and only a composition of their multiplicities
can separate one event topology from the other.
It is natural to expect that some BSM phenomena may be distinguished from the SM events
by their distinct kinematics too. For example, heavy particles can predominately decay into two other jets/particles in the central detector region, while other BSM models may ``prefer'' to populate the forward detector regions. 

We will use a simple toy consideration to calculate the number of possible
kinematic features using combinatorics with substitution.
Assume that all objects in \ncominatorics~distinct event classes populate the detector phase space according to the SM expectations. We can define a new phenomenon if two objects approach close to each other, i.e. they are the decay products of low-mass states\footnote{In this discussion, ``low-mass states'' mean BSM particles with  masses close to the sum of the masses of the daughter particles. One can also consider an alternative kinematic situation when two objects
are back-to-back in the central region of a detector if they originate from the decay products of a high-mass state. However, the production rates of such events are expected to be significantly more suppressed due to large masses, 
compared to low-mass resonances leading to collimated production of jets/particles.}. 
Such objects are still counted as two separate objects, 
but they form ensembles of kinematically unique events, and their production rate should be larger than that obtained
from pure statistical noise around the SM-defined densities. 

Let us count how many such unique kinematic topologies can exist by grouping jets and particles.
For example, consider the event topology with one jet, one $b-$jet  and one electron, such as:
$$
(1j,\> 1b,\> 1e),
$$
where we shorten the notation after removing  $0m$, $0\mu$, $0\tau$ and $0\gamma$.
This event topology creates 3 kinematically-distinct classes: 
$$
(0j,\>    1j1b,\>   1e), \quad
(1j,\>    0b, \>    1b1e), \quad 
(1j1e,\>  1b, \>    0e), 
$$
where the four-character strings, $1j1b$, $1b1e$ and $1j1e$,  represent three two-body groups
with a certain dynamic correlation between the objects in each group. 
For example,  such objects can be close to each other for a statistically
significant number of events, since they stem from exotic low-mass states. 
Experimentally, these three combinations can be viewed as invariant masses of jet+($b$-jet), $e$+($b-$jet) and jet+$e$ with associated production of other objects produced anywhere in a detector following the SM single-particle densities.
Now we can ask this question: how many such sub-classes of events exist out of \ncominatorics~total combinations?
The obtained number using numeric combinatorics is \nkinematics~(see Appendix~\ref{code}).

As before, now we need to estimate how many two-body distributions have been analyzed at the LHC.
We assume that, for each of the 3,000 event classes studied at the LHC, at least one relevant two-body kinematic distribution (such as an invariant mass) 
has been inspected, and no deviations from the SM have been found.
Therefore, for the expected \nkinematics~event classes with two-body correlation, the chances that the LHC will encounter one of these topologies, which may have an excess over the SM background, are about 2\%. 
This assumes that such events with correlations are explored uniformly across all the event topologies.  
This estimate can only be used as a conservative guide, or an upper limit on the actual LHC coverage of new phenomena, since this calculation does not consider charge topologies, correlations beyond the two-prong decays, known heavy SM particles, and other possibilities. 

For the boundary condition Eq.~(\ref{boundaryH}), which is motivated by the recent LHC studies, 
the calculated number of two-particle sub-classes
is 53,108. This leads to $\simeq 6\%$ kinematic distributions
potentially explored at the LHC.

\section{Conclusion}

The modern approach to searches for new physics at the LHC is usually based on event signatures proposed by model 
builders. It is quite clear that the LHC has good coverage of event topologies with low jet/particle multiplicities 
and hard-QCD jets. But for events with large multiplicities, where jets are treated exclusively,  the experimental coverage of the LHC is not large.

Nature can be more unpredictable, and more model-agnostic approaches can also be useful for discovering new physics in the LHC data. Our numeric analysis, guided
by the large-scale SM simulations, reveals  
that the non-observation  of new phenomena at the LHC
is not unsurprising. If a BSM signal with unusual two-particle correlations can equally be found in any of the event classes discussed in this paper, then the chance
that the LHC could detect such a new phenomenon is rather small, that is, about $2\%$ (or $6\%$), depending on the boundary condition used in the numeric calculation.
If we are only interested in jet/particle multiplicities, then the number of unexplored event classes is 81\% (55\%), leading to the probability of 19\% (45\%) for the observation of a new event topology at the LHC. 
These estimates assume that a BSM phenomenon can equally be found in any of those unexplored event classes and jets are treated
excursively. 
These values represent the upper bounds
on the probability of finding new phenomena because only
the lightest identified particles were taken into account, and the calculation
of kinematically distinct event classes includes only one feature 
(i.e. two-particle correlations). Despite the approximate nature of our calculations, they represent the first quantitative estimates obtained under the assumptions proposed in this paper. Thus, the LHC is still at the beginning of the journey to discover new physics.

In order to tackle the problem of searches for new phenomena  
in the vast number of possible event topologies  reported in this paper, novel methods of data analysis, which rely less on expectations from BSM models, should be widely used. For example, 
unsupervised machine-learning methods can automatically label
unusual event classes as anomalies. Then such anomalous events can be compared with the SM predictions. Only very recently the LHC \cite{ATLAS:2023ixc,CMS-PAS-EXO-22-026} has started its physics program of using  anomaly detection and fully unsupervised machine learning for complete event kinematics. 
For studies of multiplicities of event classes, one can train a neural network to reproduce the shapes of rates of event class as a function of their multiplicities using a small fraction of data or some control region. Comparing such shapes with the actual data would provide a useful tool to understand the ``missing information'' problem.
We hope this or similar approaches will receive the needed attention, and the general searches with machine learning will realize their full potential to discover new physics 
in the near future.

\begin{acknowledgments}
The submitted manuscript has been created by UChicago Argonne, LLC, Operator of Argonne National Laboratory (“Argonne”). Argonne, a U.S.
Department of Energy Office of Science laboratory, is operated under Contract No. DE-AC02-06CH11357. The U.S. Government retains for itself,
and others acting on its behalf, a paid-up nonexclusive, irrevocable worldwide license in said article to reproduce, prepare derivative works,
distribute copies to the public, and perform publicly and display publicly, by or on behalf of the Government.
The Department of Energy will provide public access to these results of federally sponsored research in accordance with the
DOE Public Access Plan. \url{http://energy.gov/downloads/doe-public-access-plan}. Argonne National Laboratory’s work was
funded by the U.S. Department of Energy, Office of High Energy Physics under contract DE-AC02-06CH11357.
We gratefully acknowledge the computing resources provided by the Laboratory
Computing Resource Center at Argonne National Laboratory.
\end{acknowledgments}

\appendix

\section{Code availability}
\label{code}
The \pythia\> settings, the generator code and the output data used in this study can be accessed via~\cite{BSM4LHC}. The numeric code for the combinatorial analysis is implemented in PYTHON 3  without external dependencies.

\bibliography{apssamp}

\begin{thebibliography}{15}%
\makeatletter
\providecommand \@ifxundefined [1]{%
 \@ifx{#1\undefined}
}%
\providecommand \@ifnum [1]{%
 \ifnum #1\expandafter \@firstoftwo
 \else \expandafter \@secondoftwo
 \fi
}%
\providecommand \@ifx [1]{%
 \ifx #1\expandafter \@firstoftwo
 \else \expandafter \@secondoftwo
 \fi
}%
\providecommand \natexlab [1]{#1}%
\providecommand \enquote  [1]{``#1''}%
\providecommand \bibnamefont  [1]{#1}%
\providecommand \bibfnamefont [1]{#1}%
\providecommand \citenamefont [1]{#1}%
\providecommand \href@noop [0]{\@secondoftwo}%
\providecommand \href [0]{\begingroup \@sanitize@url \@href}%
\providecommand \@href[1]{\@@startlink{#1}\@@href}%
\providecommand \@@href[1]{\endgroup#1\@@endlink}%
\providecommand \@sanitize@url [0]{\catcode `\\12\catcode `\$12\catcode `\&12\catcode `\#12\catcode `\^12\catcode `\_12\catcode `\%12\relax}%
\providecommand \@@startlink[1]{}%
\providecommand \@@endlink[0]{}%
\providecommand \url  [0]{\begingroup\@sanitize@url \@url }%
\providecommand \@url [1]{\endgroup\@href {#1}{\urlprefix }}%
\providecommand \urlprefix  [0]{URL }%
\providecommand \Eprint [0]{\href }%
\providecommand \doibase [0]{https://doi.org/}%
\providecommand \selectlanguage [0]{\@gobble}%
\providecommand \bibinfo  [0]{\@secondoftwo}%
\providecommand \bibfield  [0]{\@secondoftwo}%
\providecommand \translation [1]{[#1]}%
\providecommand \BibitemOpen [0]{}%
\providecommand \bibitemStop [0]{}%
\providecommand \bibitemNoStop [0]{.\EOS\space}%
\providecommand \EOS [0]{\spacefactor3000\relax}%
\providecommand \BibitemShut  [1]{\csname bibitem#1\endcsname}%
\let\auto@bib@innerbib\@empty
\bibitem [{\citenamefont {Sjostrand}\ \emph {et~al.}(2008)\citenamefont {Sjostrand}, \citenamefont {Mrenna},\ and\ \citenamefont {Skands}}]{Sjostrand:2007gs}%
  \BibitemOpen
  \bibfield  {author} {\bibinfo {author} {\bibfnamefont {T.}~\bibnamefont {Sjostrand}}, \bibinfo {author} {\bibfnamefont {S.}~\bibnamefont {Mrenna}},\ and\ \bibinfo {author} {\bibfnamefont {P.~Z.}\ \bibnamefont {Skands}},\ }\bibfield  {title} {\bibinfo {title} {{A Brief Introduction to PYTHIA 8.1}},\ }\href {https://doi.org/10.1016/j.cpc.2008.01.036} {\bibfield  {journal} {\bibinfo  {journal} {Comput. Phys. Commun.}\ }\textbf {\bibinfo {volume} {178}},\ \bibinfo {pages} {852} (\bibinfo {year} {2008})},\ \Eprint {https://arxiv.org/abs/0710.3820} {arXiv:0710.3820 [hep-ph]} \BibitemShut {NoStop}%
\bibitem [{\citenamefont {Chekanov}\ \emph {et~al.}(2018)\citenamefont {Chekanov}, \citenamefont {Childers}, \citenamefont {Proudfoot}, \citenamefont {Frizzell},\ and\ \citenamefont {Wang}}]{Chekanov:2017pnx}%
  \BibitemOpen
  \bibfield  {author} {\bibinfo {author} {\bibfnamefont {S.~V.}\ \bibnamefont {Chekanov}}, \bibinfo {author} {\bibfnamefont {J.~T.}\ \bibnamefont {Childers}}, \bibinfo {author} {\bibfnamefont {J.}~\bibnamefont {Proudfoot}}, \bibinfo {author} {\bibfnamefont {D.}~\bibnamefont {Frizzell}},\ and\ \bibinfo {author} {\bibfnamefont {R.}~\bibnamefont {Wang}},\ }\bibfield  {title} {\bibinfo {title} {{Precision searches in dijets at the HL-LHC and HE-LHC}},\ }\href {https://doi.org/10.1088/1748-0221/13/05/P05022} {\bibfield  {journal} {\bibinfo  {journal} {JINST}\ }\textbf {\bibinfo {volume} {13}}\bibfield  {number} {\bibinfo  {number} { (05)},\ \bibinfo {pages} {P05022}},\ }\Eprint {https://arxiv.org/abs/1710.09484} {arXiv:1710.09484 [hep-ex]} \BibitemShut {NoStop}%
\bibitem [{\citenamefont {Ball}\ \emph {et~al.}(2015)\citenamefont {Ball} \emph {et~al.}}]{Ball:2014uwa}%
  \BibitemOpen
  \bibfield  {author} {\bibinfo {author} {\bibfnamefont {R.~D.}\ \bibnamefont {Ball}} \emph {et~al.} (\bibinfo {collaboration} {NNPDF}),\ }\bibfield  {title} {\bibinfo {title} {{Parton distributions for the LHC Run II}},\ }\href {https://doi.org/10.1007/JHEP04(2015)040} {\bibfield  {journal} {\bibinfo  {journal} {JHEP}\ }\textbf {\bibinfo {volume} {04}},\ \bibinfo {pages} {040}},\ \Eprint {https://arxiv.org/abs/1410.8849} {arXiv:1410.8849 [hep-ph]} \BibitemShut {NoStop}%
\bibitem [{\citenamefont {Buckley}\ \emph {et~al.}(2015)\citenamefont {Buckley}, \citenamefont {Ferrando}, \citenamefont {Lloyd}, \citenamefont {Nordström}, \citenamefont {Page}, \citenamefont {Rüfenacht}, \citenamefont {Schönherr},\ and\ \citenamefont {Watt}}]{Buckley:2014ana}%
  \BibitemOpen
  \bibfield  {author} {\bibinfo {author} {\bibfnamefont {A.}~\bibnamefont {Buckley}}, \bibinfo {author} {\bibfnamefont {J.}~\bibnamefont {Ferrando}}, \bibinfo {author} {\bibfnamefont {S.}~\bibnamefont {Lloyd}}, \bibinfo {author} {\bibfnamefont {K.}~\bibnamefont {Nordström}}, \bibinfo {author} {\bibfnamefont {B.}~\bibnamefont {Page}}, \bibinfo {author} {\bibfnamefont {M.}~\bibnamefont {Rüfenacht}}, \bibinfo {author} {\bibfnamefont {M.}~\bibnamefont {Schönherr}},\ and\ \bibinfo {author} {\bibfnamefont {G.}~\bibnamefont {Watt}},\ }\bibfield  {title} {\bibinfo {title} {{LHAPDF6: parton density access in the LHC precision era}},\ }\href {https://doi.org/10.1140/epjc/s10052-015-3318-8} {\bibfield  {journal} {\bibinfo  {journal} {Eur. Phys. J. C}\ }\textbf {\bibinfo {volume} {75}},\ \bibinfo {pages} {132} (\bibinfo {year} {2015})},\ \Eprint {https://arxiv.org/abs/1412.7420} {arXiv:1412.7420 [hep-ph]} \BibitemShut {NoStop}%
\bibitem [{\citenamefont {Cacciari}\ \emph {et~al.}(2008)\citenamefont {Cacciari}, \citenamefont {Salam},\ and\ \citenamefont {Soyez}}]{Cacciari:2008gp}%
  \BibitemOpen
  \bibfield  {author} {\bibinfo {author} {\bibfnamefont {M.}~\bibnamefont {Cacciari}}, \bibinfo {author} {\bibfnamefont {G.~P.}\ \bibnamefont {Salam}},\ and\ \bibinfo {author} {\bibfnamefont {G.}~\bibnamefont {Soyez}},\ }\bibfield  {title} {\bibinfo {title} {{The anti-kT jet clustering algorithm}},\ }\href {https://doi.org/10.1088/1126-6708/2008/04/063} {\bibfield  {journal} {\bibinfo  {journal} {JHEP}\ }\textbf {\bibinfo {volume} {04}},\ \bibinfo {pages} {063}},\ \Eprint {https://arxiv.org/abs/0802.1189} {arXiv:0802.1189 [hep-ph]} \BibitemShut {NoStop}%
\bibitem [{\citenamefont {Cacciari}\ \emph {et~al.}(2012)\citenamefont {Cacciari}, \citenamefont {Salam},\ and\ \citenamefont {Soyez}}]{Cacciari:2011ma}%
  \BibitemOpen
  \bibfield  {author} {\bibinfo {author} {\bibfnamefont {M.}~\bibnamefont {Cacciari}}, \bibinfo {author} {\bibfnamefont {G.~P.}\ \bibnamefont {Salam}},\ and\ \bibinfo {author} {\bibfnamefont {G.}~\bibnamefont {Soyez}},\ }\bibfield  {title} {\bibinfo {title} {{FastJet User Manual}},\ }\href {https://doi.org/10.1140/epjc/s10052-012-1896-2} {\bibfield  {journal} {\bibinfo  {journal} {Eur. Phys. J.}\ }\textbf {\bibinfo {volume} {C 72}},\ \bibinfo {pages} {1896} (\bibinfo {year} {2012})},\ \bibinfo {note} {\url{http://fastjet.fr/}},\ \Eprint {https://arxiv.org/abs/1111.6097} {arXiv:1111.6097 [hep-ph]} \BibitemShut {NoStop}%
\bibitem [{\citenamefont {{ATLAS Collaboration}}(2021)}]{ATLAS:2020wgq}%
  \BibitemOpen
  \bibfield  {author} {\bibinfo {author} {\bibnamefont {{ATLAS Collaboration}}} (\bibinfo {collaboration} {ATLAS}),\ }\bibfield  {title} {\bibinfo {title} {{Search for phenomena beyond the Standard Model in events with large $b$-jet multiplicity using the ATLAS detector at the LHC}},\ }\href {https://doi.org/10.1140/epjc/s10052-020-08730-0} {\bibfield  {journal} {\bibinfo  {journal} {Eur. Phys. J. C}\ }\textbf {\bibinfo {volume} {81}},\ \bibinfo {pages} {11} (\bibinfo {year} {2021})},\ \bibinfo {note} {[Erratum: Eur.Phys.J.C 81, 249 (2021)]},\ \Eprint {https://arxiv.org/abs/2010.01015} {arXiv:2010.01015 [hep-ex]} \BibitemShut {NoStop}%
\bibitem [{\citenamefont {{CMS Collaboration}}(2019)}]{CMS:2019hsm}%
  \BibitemOpen
  \bibfield  {author} {\bibinfo {author} {\bibnamefont {{CMS Collaboration}}} (\bibinfo {collaboration} {CMS}),\ }\bibfield  {title} {\bibinfo {title} {{Search for vector-like leptons in multilepton final states in proton-proton collisions at $\sqrt{s}$ = 13 TeV}},\ }\href {https://doi.org/10.1103/PhysRevD.100.052003} {\bibfield  {journal} {\bibinfo  {journal} {Phys. Rev. D}\ }\textbf {\bibinfo {volume} {100}},\ \bibinfo {pages} {052003} (\bibinfo {year} {2019})},\ \Eprint {https://arxiv.org/abs/1905.10853} {arXiv:1905.10853 [hep-ex]} \BibitemShut {NoStop}%
\bibitem [{\citenamefont {{ATLAS Collaboration}}(2019{\natexlab{a}})}]{ATLAS:2019mra}%
  \BibitemOpen
  \bibfield  {author} {\bibinfo {author} {\bibnamefont {{ATLAS Collaboration}}} (\bibinfo {collaboration} {ATLAS}),\ }\href@noop {} {\bibinfo {title} {{Multijet simulation for 13 TeV ATLAS Analyses}}} (\bibinfo {year} {2019}{\natexlab{a}}),\ \bibinfo {note} {{ATL-PHYS-PUB-2019-017}}\BibitemShut {NoStop}%
\bibitem [{Note1()}]{Note1}%
  \BibitemOpen
  \bibinfo {note} {In our view, this number of studied event classes is very optimistic. Typically, an experimental paper has only a few such distributions, and most of them are too inclusive to pinpoint a specific event class. In addition, ATLAS and CMS often repeat measurements using similar final states. Due to the lack of exact statistics on published distributions, we prefer to consider the most optimistic scenario.}\BibitemShut {Stop}%
\bibitem [{\citenamefont {{ATLAS Collaboration}}(2019{\natexlab{b}})}]{ATLAS:2018zdn}%
  \BibitemOpen
  \bibfield  {author} {\bibinfo {author} {\bibnamefont {{ATLAS Collaboration}}} (\bibinfo {collaboration} {ATLAS}),\ }\bibfield  {title} {\bibinfo {title} {{A strategy for a general search for new phenomena using data-derived signal regions and its application within the ATLAS experiment}},\ }\href {https://doi.org/10.1140/epjc/s10052-019-6540-y} {\bibfield  {journal} {\bibinfo  {journal} {Eur. Phys. J. C}\ }\textbf {\bibinfo {volume} {79}},\ \bibinfo {pages} {120} (\bibinfo {year} {2019}{\natexlab{b}})},\ \Eprint {https://arxiv.org/abs/1807.07447} {arXiv:1807.07447 [hep-ex]} \BibitemShut {NoStop}%
\bibitem [{Note2()}]{Note2}%
  \BibitemOpen
  \bibinfo {note} {In this discussion, ``low-mass states'' mean BSM particles with masses close to the sum of the masses of the daughter particles. One can also consider an alternative kinematic situation when two objects are back-to-back in the central region of a detector if they originate from the decay products of a high-mass state. However, the production rates of such events are expected to be significantly more suppressed due to large masses, compared to low-mass resonances leading to collimated production of jets/particles.}\BibitemShut {Stop}%
\bibitem [{\citenamefont {{ATLAS Collaboration}}(2024)}]{ATLAS:2023ixc}%
  \BibitemOpen
  \bibfield  {author} {\bibinfo {author} {\bibnamefont {{ATLAS Collaboration}}} (\bibinfo {collaboration} {ATLAS}),\ }\bibfield  {title} {\bibinfo {title} {{Search for New Phenomena in Two-Body Invariant Mass Distributions Using Unsupervised Machine Learning for Anomaly Detection at s=13\,\,TeV with the ATLAS Detector}},\ }\href {https://doi.org/10.1103/PhysRevLett.132.081801} {\bibfield  {journal} {\bibinfo  {journal} {Phys. Rev. Lett.}\ }\textbf {\bibinfo {volume} {132}},\ \bibinfo {pages} {081801} (\bibinfo {year} {2024})},\ \Eprint {https://arxiv.org/abs/2307.01612} {arXiv:2307.01612 [hep-ex]} \BibitemShut {NoStop}%
\bibitem [{\citenamefont {{CMS Collaboration}}(2024)}]{CMS-PAS-EXO-22-026}%
  \BibitemOpen
  \bibfield  {author} {\bibinfo {author} {\bibnamefont {{CMS Collaboration}}} (\bibinfo {collaboration} {CMS}),\ }\href {https://cds.cern.ch/record/2892677} {\emph {\bibinfo {title} {{Model-agnostic search for dijet resonances with anomalous jet substructure in proton-proton collisions at $\sqrt{s}$ = 13 TeV}}}},\ \bibinfo {type} {Tech. Rep.}\ (\bibinfo  {institution} {CERN},\ \bibinfo {address} {Geneva},\ \bibinfo {year} {2024})\BibitemShut {NoStop}%
\bibitem [{\citenamefont {Chekanov}(2023)}]{BSM4LHC}%
  \BibitemOpen
  \bibfield  {author} {\bibinfo {author} {\bibfnamefont {S.~V.}\ \bibnamefont {Chekanov}},\ }\href@noop {} {\bibinfo {title} {{GitHub repository for the numeric code and data used in HEP-ANL-186383}}},\ \bibinfo {howpublished} {\url{https://github.com/chekanov/HEP-ANL-186383}} (\bibinfo {year} {2023})\BibitemShut {NoStop}%
\end{thebibliography}%

\end{document}